\documentclass[journal]{IEEEtran}
\usepackage{bm}
\usepackage{graphicx}
\usepackage{physics}
\usepackage{multirow}
\usepackage{makecell}
\usepackage{hyperref}
\usepackage{float}
\usepackage{tikz}
\usepackage{amsmath}
\usepackage{amssymb}
\ifCLASSOPTIONcompsoc
\usepackage[caption=false,font=normalsize,labelfont=sf,textfont=sf]{subfig}
\else
\usepackage[caption=false,font=footnotesize]{subfig}
\fi

\newcommand{\comment}[1]{{{#1}}}
\renewcommand{\eqref}[1]{{equation (\ref{#1})}}

\begin{document}
	
\title{Tracking Single Particles using Surface Plasmon Leakage Radiation Speckle}

\author{Joel Berk, Carl Paterson and Matthew R. Foreman
	\thanks{This work was funded by the Engineering and Physical Sciences Research Council (EPSRC) (1992728) and the Royal Society.}%
	\thanks{Joel Berk, Carl Paterson and Matthew R. Foreman are with Blackett Laboratory, Imperial College London, Prince Consort Road, London, SW7 2BW (email: matthew.foreman@imperial.ac.uk)}
}

\maketitle

\begin{abstract}
	Label free tracking of small bio-particles such as proteins or viruses is of great utility in the study of biological processes, however such experiments are frequently hindered by weak signal strengths and a susceptibility to scattering impurities. To overcome these problems we here propose a novel technique leveraging the enhanced sensitivity of both interferometric detection and the strong field confinement of surface plasmons. Specifically, we show that interference between the field scattered by an analyte particle and a  speckle reference field, derived from random scattering of surface plasmons propagating on a rough metal film, enables particle tracking with sub-wavelength accuracy. We present the analytic framework of our technique and verify its robustness to noise through Monte Carlo simulations.
\end{abstract}
	
\section{Introduction}
\IEEEPARstart{L}{ocalising} and tracking small biological particles, such as viruses or proteins, has played an important part in enabling advances in our understanding of biological processes at the microscopic and nanoscopic level \cite{McDonald2018VisualizingSensitivity,Kukura2009High-speedVirus,Giepmans2006TheFunction}. 
Such studies require an ability to detect, monitor and analyse processes dynamically without the usual ensemble averaging inherent in many techniques. Optical microscopy has played a pivotal role, however small biological particles typically interact weakly with light thereby requiring detection of faint signals which are prone to the effects of noise. Labelling analyte particles, for example with a fluorophore \cite{Giepmans2006TheFunction}, is a proven technique for increasing signal strength and has thus enabled not only single molecule detection \cite{Moerner1989,Orrit1990}, but also prompted development of a wealth of super-resolution microscopy techniques.  Photo-activated localisation microscopy (PALM) and stochastic optical reconstruction microscopy (STORM), for instance, are capable of localising individual biomolecules with nanometre precision \cite{EricBetzig2006ImagingResolution,Rust2006Sub-diffraction-limitSTORM}. 

Although analyte labelling can help mitigate the effects of noise, label free techniques are often preferable in a biological context so as to avoid modifying the dynamics and function of analyte particles \cite{Ellen2012,Toseland2013FluorescentProteins}. Label-free single particle detection methods have thus attracted significant research attention in recent years. Primarily, such techniques seek to improve sensitivity through modified detection schemes or to enhance light-matter interactions via strong field confinement. Interferometric scattering microscopy (iSCAT), for example, leverages interference between light scattered by an analyte particle and a known reference field so as to significantly increase the magnitude of the relevant signal. Detection and tracking of single unlabelled proteins and viruses using iSCAT has thus recently been reported \cite{Kukura2009High-speedVirus,Taylor2019InterferometricScattering,Liebel2017UltrasensitiveProteins,Cole2017Label-FreeMicroscopy}. Alternatively, resonance based sensing techniques typically monitor analyte induced pertubations to a measured resonance line profile, such as frequency shifts or mode broadening. Sensitivity in these systems is ultimately limited by the resonance linewidth since this dictates the size of detectable changes in the presence of noise \cite{Piliarik:09,Foreman2014a}. High $Q$-factor resonances, for example whispering gallery modes \cite{Vollmer2012Label-freeDevices,Foreman2015WhisperingSensors} or photonic crystal cavity modes \cite{Konopsky2018PhotonicPeak}, have thus been used to achieve single protein detection  \cite{Vollmer2012Label-freeDevices} and characterisation of single nanoparticles \cite{Zhu2011,Foreman2017}. In contrast, plasmonic resonances are relatively broad due to losses inherent in metals, however, they can benefit from particularly strong field confinement. Accordingly, surface plasmon resonances (SPRs) in thin metallic films have been used extensively for sensing of biomolecules in bulk \cite{Homola1999SurfaceReview,Lundstrom1994Real-timeAnalysis}, whereas detection of single particle binding events and conformational changes has been achieved using localised surface plasmon resonance (LSPR) in metallic nanoparticles \cite{JeffreyN.Anker2008BiosensingNanosensors,Zijlstra2012OpticalNanorod}. Strong near fields associated with plasmonic modes can also be leveraged for surface enhanced Raman spectroscopy (SERS) \cite{Laing2017Surface-enhancedBiosensing,Rieder1997ProbingScattering,Wang2013Surface-enhancedBiology}, in turn enabling identification of analyte components. Hybrid photonic-plasmonic resonance based systems have also been shown to enable high sensitivity particle detection \cite{DeAngelis2008,Baaske2014}. 

Whilst resonance based modalities have a proven record for sensitive detection, they provide only very limited positional information about analyte particles and are thus less suitable for localisation and tracking based applications. Tracking of biological molecules is currently achieved primarily through imaging, for example using multifocal plane microscopy \cite{Toprak2007, Ram2012}, dynamic localisation \cite{Hou:17} or holographic imaging \cite{Memmolo:15}. Typically, a series of images of the analyte is taken and subsequent image analysis used to extract the particle trajectory \cite{Saxton1997SINGLE-PARTICLEDynamics,Carter2005TrackingEvaluation,Shen2017SingleApplications}. The accuracy of such a method thus depends on both the algorithm and imaging modality. Super-resolution methods using fluorescent labels can achieve nanometre scale tracking precision \cite{Shen2017SingleApplications}, while iSCAT can similarly achieve errors on the order 2~nm \cite{Kukura2009High-speedVirus}.

\begin{figure*}[t]
	\centering
	\includegraphics[width=0.9\textwidth]{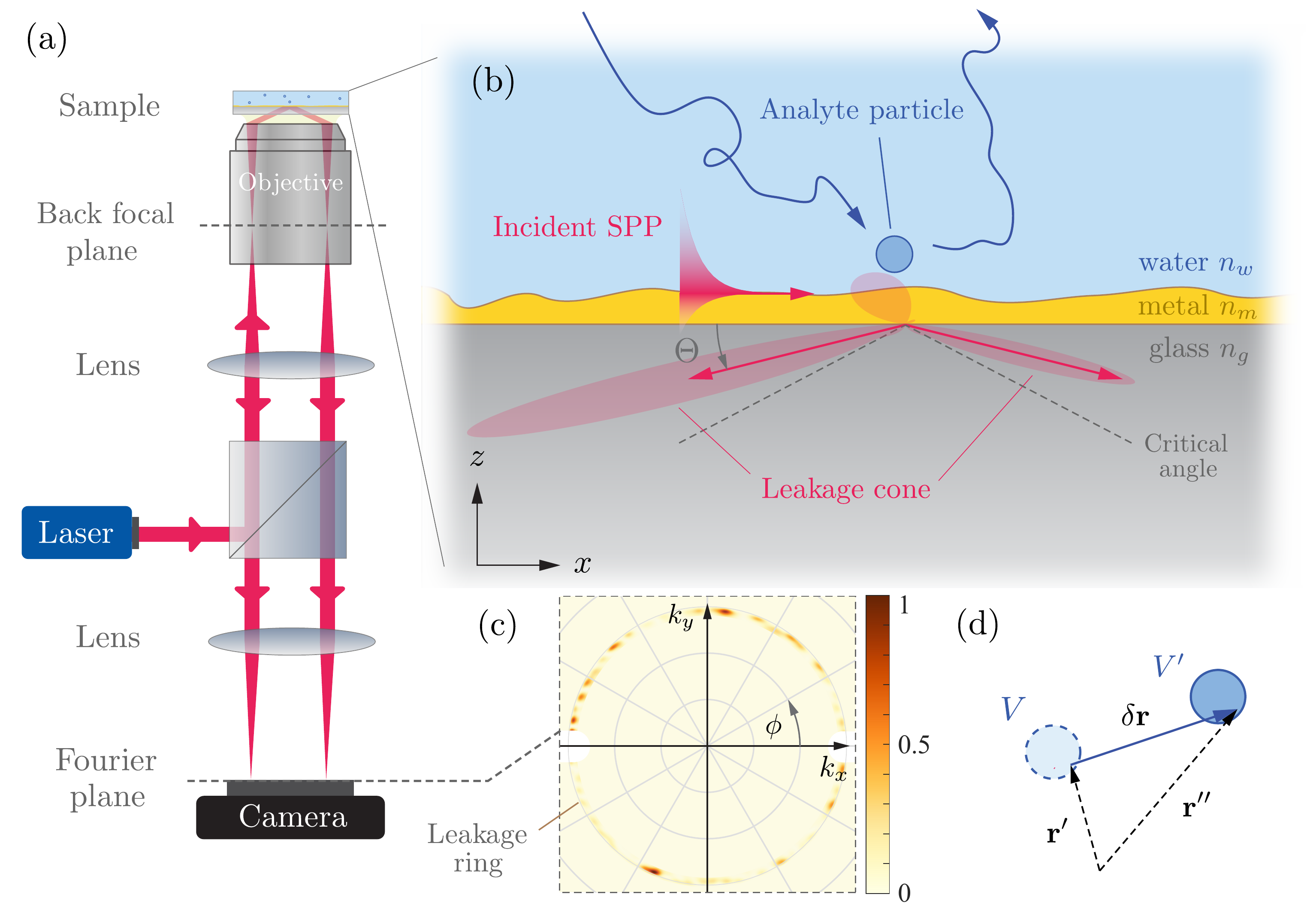}
	\caption{(a) A schematic of the proposed experimental setup for particle tracking based on Kretschmann coupling and Fourier plane imaging of leakage radiation. (b) Thin film structure, supporting an incident plane SPP wave which propagates along the metal-dielectric interface and scatters from an analyte particle in the neighbouring aqueous solution. The scattering pattern from a dipole, with dipole moment proportional to the incident SPP field, is overlayed (red) demonstrating the highly directional scattering into the leakage cone. (c) Random scattering from a rough metal thin film gives rise to a ring-like speckle pattern in the Fourier plane. (d) Coordinate shift used to describe translation of the particle.}  
	\label{fig:exp_setup}
\end{figure*}

In this work we propose a technique for tracking small biological particles which utilises both the enhanced sensitivity of interferometric detection akin to iSCAT, whilst also benefiting from enhanced local fields inherent to use of plasmonic resonances. Use of surface plasmons in iSCAT has been shown to be a sensitive technique for imaging of single exosomes \cite{Yang2018InterferometricExosomes}, however, due to its coherent nature, iSCAT is particularly susceptible to scattering impurities whereby optical speckle can degrade image quality \cite{Taylor2019InterferometricScattering}. This poses a significant obstacle for tracking particles in complex scattering scenarios, such as lipid membranes or cellular environments.
Our proposed tracking technique utilises a speckle based reference signal formed from random scattering of surface plasmon polaritons (SPPs) \cite{Foreman2019a} to help overcome such limitations and hence move towards study of biomolecular function in more realistic biological settings. \comment{We foresee the technique being able to study biological processes occuring at surfaces such as: binding of proteins to receptors on a membrane, for example the interaction of epidermal growth factor (EGF) with epidermal growth factor receptor (EGFR) \cite{Kim17}; the motion of molecular motors such as myosin along actin filaments \cite{myosin}; and the movement of proteins such as EGFR on or through membranes \cite{membrane_transport,egfr_membrane,Sandoghdar14}. Moreover, DNA origami structures offer nanometre level control of the local environment, offering a route to molecular scale machinery or assembly lines \cite{DNA_origami}. Studying the interaction of molecules with these structures functionalised on a surface can provide valuable information towards developing such molecular machines \cite{Jungmann10}.} We begin in Section~\ref{sec:theory} by introducing the proposed system and discussing the origin of the random reference signal. Section~\ref{sec:tracking} builds on the theoretical scattering model introduced in Section~\ref{sec:theory} to derive the proposed tracking algorithm, which is subsequently verified in Section~\ref{sec:results} by means of numerical simulation. Algorithm performance in the presence of noise is also studied. Finally, a detailed critical discussion of our proposed method is presented in Section~\ref{sec:discussion} before our conclusions are given in Section~\ref{sec:conclusion}.

\section{Theoretical scattering model}\label{sec:theory}

We consider a thin film geometry, as depicted in Fig. \ref{fig:exp_setup}, consisting of a thin metallic layer (with electric permittivity $\epsilon_m$) depositied on a glass substrate (electric permittivity $\epsilon_g$), immersed in an aqueous solution (electric permittivity $\epsilon_w$) containing the analyte particles. Initially assuming the interfaces are optically smooth, the electric field, $\mathbf{E}_0$, of a plane SPP field at a position $\mathbf{r}=(x,y,z)$ in the upper half-space ($z>0$) at time $t$, propagating in the $x$-direction with frequency $\omega$, can be written in the form
\begin{equation}
\mathbf{E}_0(\mathbf{r},t)=
A\begin{pmatrix}
i\kappa_w \\
0 \\
-k_{sp}
\end{pmatrix}\exp(ik_{sp}x-\kappa_w z-i\omega t),
\label{eq:SPP_def_E}
\end{equation}
where $A$ dictates the amplitude of the SPP at the surface, $k_{sp}$ is the SPP wavenumber, $\kappa_w$ is the out of plane decay constant and $k_{sp}^2-\kappa_w^2=\epsilon_w\omega^2/c^2$ \cite{Zayats2005,Maier2007}. Metallic losses imply that  $k_{sp}=k_{sp}'+ik_{sp}''$ and $\kappa_w=\kappa'_w+i\kappa_w''$ are complex, whereby we denote the real and imaginary parts using single and double prime notation respectively. Notably,   the dielectric-metal-dielectric structure shown in Fig.~\ref{fig:exp_setup} can support two distinct SPP modes due to coupling between the interfaces, namely a long range or short range mode \cite{Zayats2005}. Although the values of $k_{sp}$ and $\kappa_w$ differ between these modes, the field in the upper half space is nevertheless of the form of \eqref{eq:SPP_def_E}. In general, the dispersion relation of the SPP modes means $k_{sp}'>\sqrt{\epsilon_w}k_0$, where $k_0=\omega/c$ is the free space wavenumber. This condition prevents coupling between SPP modes and propagating optical modes in the upper space \cite{Zayats2005}. On the other hand, the photon wavenumber in the glass substrate can be designed to be larger than the SPP wavenumber of the short range SPP, i.e. $n_gk_0>k_{sp}'$ ($n_g = \sqrt{\epsilon_g}$ is the refractive index of the substrate), therefore allowing for excitation using a Kretschmann type configuration as shown in Fig.~\ref{fig:exp_setup}(a), in addition to leakage of short range SPPs into outgoing optical waves, commonly known as leakage radiation \cite{Drezet2008LeakagePolaritons}. Specifically, the transverse momentum of the leakage radiation matches that of the SPP, such that the outgoing optical wave propagates in the forward direction at an angle $\Theta$ relative to the surface normal where $n_gk_0\sin\Theta=k_{sp}'$. We note that although Fig.~\ref{fig:exp_setup}(a) depicts a Kretschmann coupling setup \cite{Raether1988}, our proposed technique works equally well with other SPP excitation schemes e.g. grating or end fire configurations~\cite{Raether1988,Stegeman:83}.

Practically the metallic film will not be perfectly smooth, either due to fabrication imperfections or intentional design, which in turn gives rise to scattering of SPPs. Predominantly, scattered SPPs couple to secondary SPPs propagating in random directions \cite{SoNdergaard}. Through coupling into short range SPPs and their subsequent leakage, a cone of radiation is hence observed in the lower half space, where the cone half angle is again defined by $\Theta$. Direct diffuse scattering into propagating optical modes in either half space can also occur, however, this process is typically weak (as indicated by the dipole scattering pattern shown in Fig.~\ref{fig:exp_setup}(b)). Due to the random nature of the surface roughness and scattering of plasmons, the resulting leakage field is formed from the random interference of many coherent wavefronts and thus  exhibits a granular pattern of bright and dark spots, i.e. speckle.  This speckle field shall be denoted $\mathbf{E}_b(\mathbf{r})$, and is notably a fixed function of $\mathbf{r}$ for a given surface profile and coupling scheme. The conical scattering is most easily observed through Fourier plane imaging \cite{Dominguez2014}, as depicted in Fig.~\ref{fig:exp_setup}(a), whereby a speckle ring is evident (see Fig.~\ref{fig:exp_setup}(c)). The speckle field in the Fourier plane will be denoted $\widetilde{\mathbf{E}}_b(\mathbf{k}_{\parallel})$, where $\mathbf{k}_{\parallel} =(k_x,k_y)$ denotes the transverse wavevector of each angular component of the scattered field. \comment{In addition to elastic scattering of the SPP, the surface roughness modifies the SPP dispersion relation \cite{Maradudin1976}, producing a shift in the resonance wavenumber and increased attenuation relative to the smooth interface case. For this work, however, the functional dependence of $k_{sp}$ on the thin film structure and surface roughness need not be considered, since it only dictates the opening angle and width of the ring of leakage radiation which can instead be treated as experimentally determined parameters.}

When an analyte particle is sufficiently close to the surface, in addition to the scattering processes discussed above an incident SPP can also scatter directly from the analyte particle. Assuming such scattering is sufficiently weak, as is relevant for small biological particles, we can treat the additional scattering using a single scattering approximation whereby the total scattered field is given by $\mathbf{E}(\mathbf{r})=\mathbf{E}_b(\mathbf{r})+\mathbf{E}_s(\mathbf{r})$, where $\mathbf{E}_s(\mathbf{r})$ is the field scattered from the analyte. The scattered intensity at any point is thus $I(\mathbf{r})=\abs{\mathbf{E}_b(\mathbf{r})+\mathbf{E}_s(\mathbf{r})}^2=I_b(\mathbf{r})+I_s(\mathbf{r})+2\Re(\mathbf{E}_b^*(\mathbf{r})\cdot\mathbf{E}_s(\mathbf{r}))$ (neglecting prefactors), where $I_b(\mathbf{r}) = \abs{\mathbf{E}_b(\mathbf{r})}^2$ denotes the intensity of the background speckle and $I_s(\mathbf{r}) = \abs{\mathbf{E}_s(\mathbf{r})}^2$ is the intensity scattered by the analyte particle.  Equivalent expressions also hold in the Fourier plane.  In addition to the two direct scattering intensities, we note that the total scattered field also includes a term describing interference between the background leakage speckle field and that scattered from the analyte particle. It is this interference term which forms the basis of our proposed tracking technique. 

\subsection{Translation of particle}
To be able to track the motion of an analyte particle, we must first consider how an observed signal changes upon particle translation. Therefore, we consider the scattered electric field $\mathbf{E}_s(\mathbf{r};\mathbf{r}_p)$ for a particle occupying a volume $V$ centred on $\mathbf{r}_p$.
 Within the Born approximation \cite{Akkermans2007MesoscopicPhotons,Mishchenko2006MultipleBackscattering} the scattered field can be expressed
\begin{equation}\label{eq:born}
    \mathbf{E}_s(\mathbf{r};\mathbf{r}_p)=\frac{\omega^2}{c^2}\int_{\mathbf{r'}\in V}\left[\epsilon_p(\mathbf{r'})-\epsilon_w\right] G(\mathbf{r},\mathbf{r'})\cdot\mathbf{E}_0(\mathbf{r'})d^3\mathbf{r'}
\end{equation}
where the dielectric function of the particle, $\epsilon_p(\mathbf{r})$, is allowed to take an arbitrary form  within the particle and $G(\mathbf{r},\mathbf{r'})$ is the Green's tensor for the thin film geometry which describes the scattering pattern for a dipolar source, \comment{including the effects of the thin film structure on the dipolar emission}. Small (i.e. sub wavelength) particles in particular give rise to a dipole scattering pattern, a cross-section of which is shown in Fig.~\ref{fig:exp_setup}(b). As discussed above, when placed close to a metal film, scattering from a dipole couples strongly into SPP modes which subsequently leak into the leakage radiation cone, as shown by the two narrow angular lobes in the scattering pattern.

When the particle is shifted by $\delta\mathbf{r}=(\delta x,\delta y,\delta z)$, the scattered field can be similarly calculated by integrating over a volume $V'$, translated by $\delta\mathbf{r}$ (see Fig. \ref{fig:exp_setup}(d)), with a translated dielectric function $\epsilon_p'(\mathbf{r})$. Assuming the particle is spherically symmetric or that the particle is so small that internal variation of $\epsilon_p$ can be neglected, the transformed and original dielectric functions are related according to $\epsilon'_p(\mathbf{r}+\delta\mathbf{r})=\epsilon_p(\mathbf{r})$. Correspondingly, the scattered field from the shifted particle, $\mathbf{E}_s(\mathbf{r};\mathbf{r}_p+\delta\mathbf{r})$, is given by
\begin{equation}\label{eq:translate_field_interim}
        \mathbf{E}_s(\mathbf{r};\mathbf{r}_p+\delta\mathbf{r})=\frac{\omega^2}{c^2}\int_{\mathbf{r''}\in V'}[\epsilon_p'(\mathbf{r''})-\epsilon_w] G(\mathbf{r},\mathbf{r''})\mathbf{E}_0(\mathbf{r''})d^3\mathbf{r''}.
\end{equation}
Upon making the change of variables $\mathbf{r''}=\mathbf{r'}+\delta\mathbf{r}$, the integral becomes
\begin{align}\label{eq:translate_field_1}
        &\mathbf{E}_s(\mathbf{r};\mathbf{r}_p+\delta\mathbf{r})=\nonumber\\
        &\quad \frac{\omega^2}{c^2}\int_{\mathbf{r'}\in V}[\epsilon_p(\mathbf{r'})-\epsilon_w]G(\mathbf{r},\mathbf{r'}+\delta\mathbf{r})\mathbf{E}_0(\mathbf{r'}+\delta\mathbf{r})d^3\mathbf{r'}.
\end{align}
Furthermore, it follows from \eqref{eq:SPP_def_E} that  $\mathbf{E}_0(\mathbf{r'}+\delta\mathbf{r})=e^{ik_{sp}\delta x}e^{-\kappa_w\delta z}\mathbf{E}_0(\mathbf{r'})$. 
The scattered field in the Fourier plane, $\widetilde{\mathbf{E}}_s(\mathbf{k}_\parallel;\mathbf{r}_p)$,  follows by evaluting the 2D Fourier transform of \eqref{eq:translate_field_1} evaluated at the glass-metal interface, taken with respect to the transverse position vector $\boldsymbol{\rho}=(x,y)$. Notably, since the Green's tensor $G(\mathbf{r},\mathbf{r}'') = G(\boldsymbol{\rho}-\boldsymbol{\rho}'',z,z'')$ is a function of the difference of the transverse position vectors $\boldsymbol{\rho}-\boldsymbol{\rho}''$, the translated Fourier space Green's tensor, $\widetilde{G}$ can be easily related to the unshifted Fourier space Green's tensor  \cite{Novotny2012PrinciplesNano-optics} through
\begin{equation}\label{eq:translate_G0}
\widetilde{G}(\mathbf{k}_\parallel;\mathbf{\rho'}+\delta\boldsymbol{\rho},z,z'+\delta z)=\widetilde{G}(\mathbf{k}_\parallel;\mathbf{\rho'},z,z')e^{i\mathbf{k}_\parallel\cdot\delta\boldsymbol{\rho}} e^{ik_z\delta z},
\end{equation}
where $k_z=(\epsilon_wk_0^2-k_\parallel^2)^{1/2}$ and $\delta\boldsymbol{\rho}=(\delta x, \delta y)$ is the transverse component of the shift. Substituting \eqref{eq:translate_G0} into the Fourier transform of \eqref{eq:translate_field_1}, the field scattered by the translated particle in the far field in the direction defined by $\mathbf{k}_\parallel$ is hence found to be
\begin{equation}\label{eq:E_dr_E_0_general}
    \widetilde{\mathbf{E}}_s(\mathbf{k}_\parallel;\mathbf{r}_p+\delta\mathbf{r})=e^{ik_{sp}\delta x}e^{-\kappa_w\delta z}e^{i\mathbf{k}_\parallel\cdot\delta\boldsymbol{\rho}}e^{ik_z\delta z}\widetilde{\mathbf{E}}_s(\mathbf{k}_\parallel;\mathbf{r}_p).
\end{equation}
Since the scattered $\widetilde{\mathbf{E}}_s$ in the glass substrate is strongly confined to the leakage radiation ring, we henceforth consider the case where $\mathbf{k}=n_gk_0(\cos\phi\sin\Theta,\sin\phi\sin\Theta,\cos\Theta)$. From \eqref{eq:E_dr_E_0_general}, it follows that the scattered fields at azimuthal coordinate $\phi$ on the leakage radiation ring for a shifted and unshifted particle are related by
\begin{equation}\label{eq:E_dr_E_0_ring}
     \widetilde{\mathbf{E}}_s(\phi;\mathbf{r}_p+\delta\mathbf{r})=e^{i\Psi(\phi;\delta \boldsymbol{\rho},\delta z)}e^{-\Lambda(\delta x,\delta z)}\widetilde{\mathbf{E}}_s(\phi;\mathbf{r}_p),
\end{equation}
where we have introduced the phase shift and decay functions $\Psi(\phi;\delta\boldsymbol{\rho}, \delta z)$ and $\Lambda(\delta x,\delta z)$ defined as
\begin{align}
     \Psi(\phi;\delta\boldsymbol{\rho},\delta z)&=k_{sp}'\delta x(1+\cos\phi)+k_{sp}'\delta y\sin\phi-\kappa_w''\delta z\label{eq:phase_shift}\\
      \Lambda(\delta x,\delta z)&=k_{sp}''\delta x+\left[\kappa_w'+\left(k_{sp}'^2-\epsilon_wk_0^2\right)^{1/2}\right]\delta z.
\end{align}
Note that since the SPP wavenumber is larger than the wavenumber in the upper dielectric (aqueous solution), $k_z=i(k_{sp}'^2-\epsilon_wk_0^2)^{1/2}$ is imaginary and thus the $\exp(ik_z\delta z)$ factor in \eqref{eq:E_dr_E_0_general} represents a decay factor. From \eqref{eq:E_dr_E_0_ring} we thus see that as the analyte particle moves the direct scattered field in the leakage ring acquires an additional phase shift $\Psi$ with respect to the background speckle reference field, in addition to a change in amplitude. Accordingly the total field $\widetilde{\mathbf{E}}$ changes in a predictable manner, enabling the shift of the particle to be determined as we now discuss.

\section{Tracking algorithm}\label{sec:tracking}

\begin{figure*}
	\centering
	\includegraphics[width=0.9\textwidth]{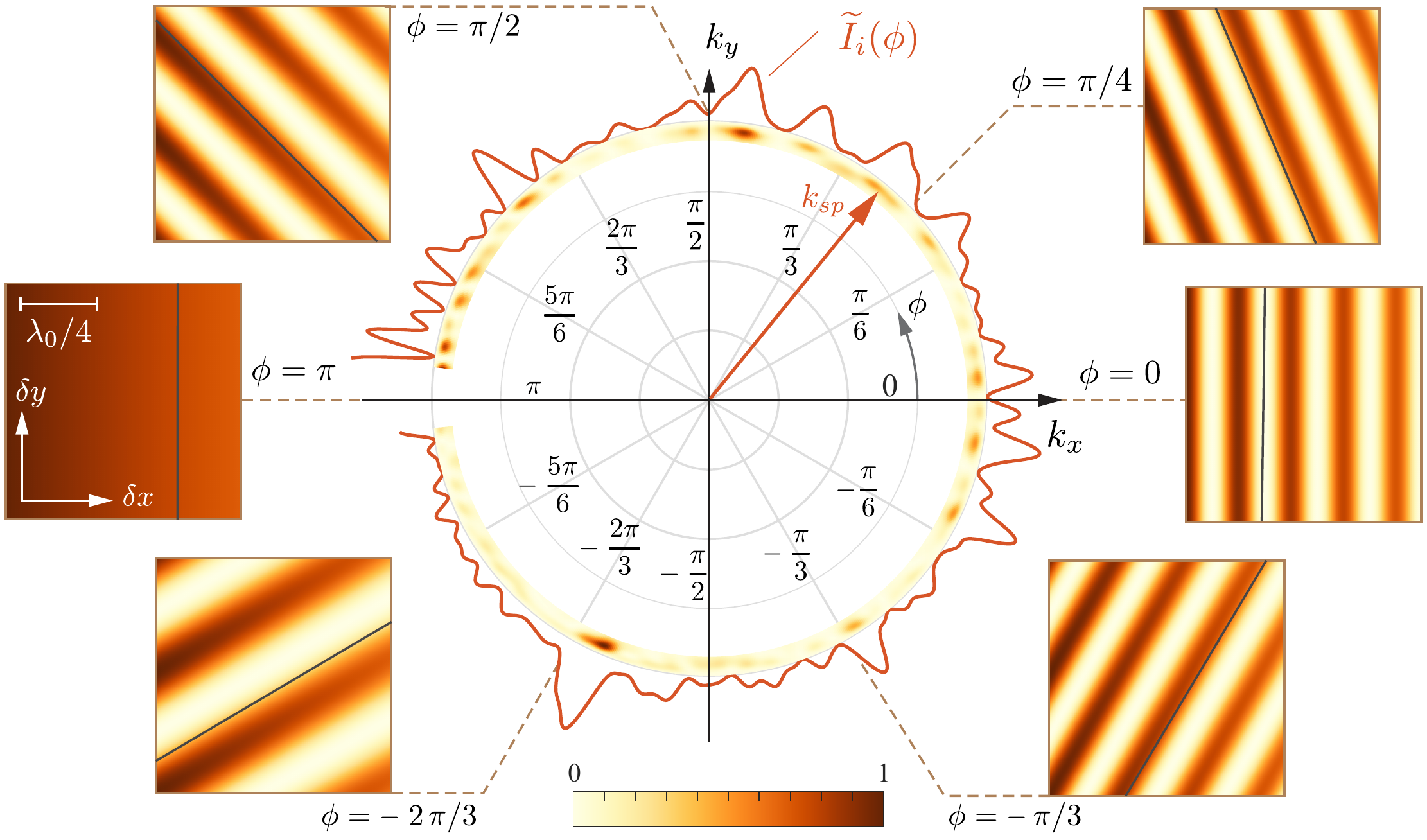}
	\caption{(center) Example speckle intensity in the Fourier plane and corresponding cross-section $\widetilde{I}_i(\phi)$ taken at $\theta = \Theta$. Note only the scattered speckle intensity has been plotted.  (panels) Fringe patterns one would see at different points on the ring from scanning the analyte particle in the $x$ and $y$ directions. In the $\phi=\pi$ direction, no fringes are seen as the phase of $\mathbf{E}_s$ does not change, so only decay effects are seen. At each $\phi$, the fringe pattern has an unknown offset. as depicted by the solid black lines, due to the random phase of the background speckle.}
	\label{fig:fringe_patterns}
\end{figure*}

In a tracking scenario it is necessary to collect multiple frames of data over time so as to capture the dynamics of the analyte particle. In our case, we consider a (small) particle moving near the surface of the metal film, such that during the $i$th frame ($i=1,2,\ldots$) the particle is located at $\mathbf{r}_i$ and the corresponding intensity distribution in the pupil plane is recorded, specifically the leakage radiation ring which we denote $\widetilde{I}_i(\phi)$. Note that we assume the time resolution of measurements is sufficient to be able to negate the effects of motion blurring. Under the assumption the particle is weakly scattering, we may assume $I_s\ll I_b$, allowing $\widetilde{I}_i$ to be expressed
\begin{equation}\label{eq:tot_intensity}
        \widetilde{I}_i(\phi)\approx \widetilde{I}_b(\phi)+2\Re\left[\widetilde{\mathbf{E}}_b^*(\phi)\cdot\widetilde{\mathbf{E}}_s(\phi;\mathbf{r}_i)\right]
\end{equation}
where $\widetilde{I}_b$ and $\widetilde{I}_s$ are defined analogously to above. The interference term depends on particle position, tracing out different fringe patterns at different points, $\phi$, on the ring (see Fig.~\ref{fig:fringe_patterns}). It also depends on the phase of the background speckle, which is unknown and varies randomly. As a result, it is not possible to extract the shift using just two frames since the fringe displacement is consequently also different and unknown for each $\phi$. To overcome this issue, our tracking method simultaneously considers three frames (somewhat analogous to phase shifting interferometry \cite{Bruning1974DigitalLenses,Lai1991GeneralizedInterferometry}, in which three or more known reference phases are used to calculate an unknown wavefront phase). It is assumed that a reference measurement of $\widetilde{I}_b(\phi)$, which is fixed for a given surface roughness profile, is taken before  the analyte particle moves into the sensing volume. Importantly, we note that we do not need to measure the phase of the reference speckle pattern. Taking three data frames (labelling them as $i=1,2,3$), we can write the intensity profiles as
\begin{align}
     \Delta \widetilde{I}_{1}(\phi)&=2\Re\left[\widetilde{\mathbf{E}}_b^*(\phi)\cdot\widetilde{\mathbf{E}}_s(\phi;\mathbf{r}_2)e^{i\Psi_{21}}e^{-\Lambda_{21}}\right]\label{eq:frame1}\\
     \Delta \widetilde{I}_{2}(\phi)&=2\Re\left[\widetilde{\mathbf{E}}_b^*(\phi)\cdot\widetilde{\mathbf{E}}_s(\phi;\mathbf{r}_2)\right]\label{eq:frame2}\\
     \Delta \widetilde{I}_{3}(\phi)&=2\Re\left[\widetilde{\mathbf{E}}_b^*(\phi)\cdot\widetilde{\mathbf{E}}_s(\phi;\mathbf{r}_2)e^{i\Psi_{23}}e^{-\Lambda_{23}}\right],\label{eq:frame3}
\end{align}
where we have defined $\Delta \widetilde{I}_i(\phi) = \widetilde{I}_i(\phi) - \widetilde{I}_b(\phi)$, \eqref{eq:E_dr_E_0_ring} has been used to express $\widetilde{\mathbf{E}}_s(\phi;\mathbf{r}_{1})$ and $\widetilde{\mathbf{E}}_s(\phi;\mathbf{r}_{3})$ in terms of  $\widetilde{\mathbf{E}}_s(\phi,\mathbf{r}_{2})$, $\Psi_{ij}$ and $\Lambda_{ij}$ are $\Psi(\phi;\delta\mathbf{r}_{ij})$ and $\Lambda(\delta\mathbf{r}_{ij})$ respectively, and $\delta\mathbf{r}_{ij}=\mathbf{r}_i-\mathbf{r}_j$. Expanding the complex exponential in \eqref{eq:frame1} into real and imaginary parts gives
\begin{equation}\label{eq:frame1_exp}
    \Delta \widetilde{I}_{1}(\phi)=e^{-\Lambda_{21}}\cos\Psi_{21}\Delta \widetilde{I}_2(\phi)-e^{-\Lambda_{21}}\sin\Psi_{21}\widetilde{K}(\phi),
\end{equation}
where we have also used \eqref{eq:frame2} and defined $\widetilde{K}(\phi)=2\Im[\widetilde{\mathbf{E}}_b^*(\phi)\cdot\widetilde{\mathbf{E}}_s(\phi;\mathbf{r}_2)]$. Following the same process with \eqref{eq:frame3} results in
\begin{equation}\label{eq:frame3_exp}
    \Delta \widetilde{I}_{3}(\phi)=e^{-\Lambda_{23}}\cos\Psi_{23}\Delta \widetilde{I}_2(\phi)-e^{-\Lambda_{23}}\sin\Psi_{23}\widetilde{K}(\phi).
\end{equation}
Multiplying \eqref{eq:frame1_exp} by $e^{-\Lambda_{23}}\sin\Psi_{23}$ and \eqref{eq:frame3_exp} by $e^{-\Lambda_{21}}\sin\Psi_{21}$ and subtracting to eliminate $\widetilde{K}(\phi)$ gives a single equation for each point $\phi$ on the ring, dependent only on the three measured intensity differences and the shifts in position between each frame, specifically
\begin{equation}\label{eq:cost_func}
\mathbf{u}\cdot\mathbf{\Delta}=0
\end{equation}
where 
\begin{align*}
\mathbf{u}&=\begin{pmatrix}
e^{-\Lambda_{23}}\sin\Psi_{23} \\
e^{-\Lambda_{23}-\Lambda_{21}}\sin(\Psi_{21}-\Psi_{23})\\
-e^{-\Lambda_{21}}\sin\Psi_{21}
\end{pmatrix}
\end{align*} 
and $\mathbf{\Delta}=( \Delta \widetilde{I}_1, \Delta \widetilde{I}_2, \Delta \widetilde{I}_3)^T$.
Note that in arriving at  \eqref{eq:cost_func} we have used the trigonometric identity $\cos\Psi_{23}\sin\Psi_{21}-\cos\Psi_{21}\sin\Psi_{23}=\sin(\Psi_{21}-\Psi_{23})$. Equation~(\ref{eq:cost_func}) holds for any value of $\phi$ on the ring. Assuming then that the intensity is sampled at $N$ discrete angles $\phi_k$, $k=1,2,\ldots,N$ we can form a set of $N$ equations with $6$ unknowns (the three components of $\delta\mathbf{r}_{21}$ and $\delta\mathbf{r}_{23}$).

To find the the shifts in particle position, a least squares solution can be used, with the estimates of the steps given by the solution to the 6D (global) minimisation
\begin{equation}
(\delta\mathbf{\hat{r}}_{21},\delta\mathbf{\hat{r}}_{23})=\underset{(\delta\mathbf{r}_{21},\delta\mathbf{r}_{23})}{\arg\min}\sum_{k=1}^{N}\left(\frac{\mathbf{u}(\phi_k)\cdot\mathbf{\Delta}(\phi_k)}{\abs{\mathbf{u}(\phi_k)}}\right)^2,
\end{equation}
where $\delta\mathbf{\hat{r}}_{ij}$ denotes an estimate of the true step $\delta\mathbf{r}_{ij}$. The vector $\mathbf{u}$ has been normalised by its magnitude in order to exclude the $\mathbf{u}=\mathbf{0}$ solution, corresponding to no shift. By applying this algorithm to sets of 3 frames, the trajectory of a particle from frame to frame over an arbitrary number of frames may be reconstructed.
\subsection{Trajectory Consistency Check}
Since the algorithm uses three frames to reconstruct a pair of consecutive steps simultaneously, it can provide two estimates for the same step generated from different measurements. For example, $\delta\mathbf{r}_{23}$ can be estimated from frames 1,2,3 or from frames 2,3,4, where each estimate arises from the minimisation of a different function. Comparing the two estimates of any given step can thus provide a consistency check of the retrieved trajectory, with the two estimates required to agree within some small margin for error. In particular, inconsistencies can occur if either the minimisation procedure fails to find the global minimum, or the effects of noise or a non-negligible direct scattering term  mean that the true step  no longer corresponds to the global minimum.  In the first case, the minimisation procedure can be modified or rerun for the two relevant sets of frames until consistent global minima for each are found. For example, for a multi-start minimisation algorithm \cite{Marti2003}, additional random starting points can be used. In the second case,  although all global minimisation algorithms will not yield the correct step, the consistency check nevertheless provides an indication that a particular step is incorrect.

\begin{figure*}[t]
	\centering
	\includegraphics[width=0.85\textwidth]{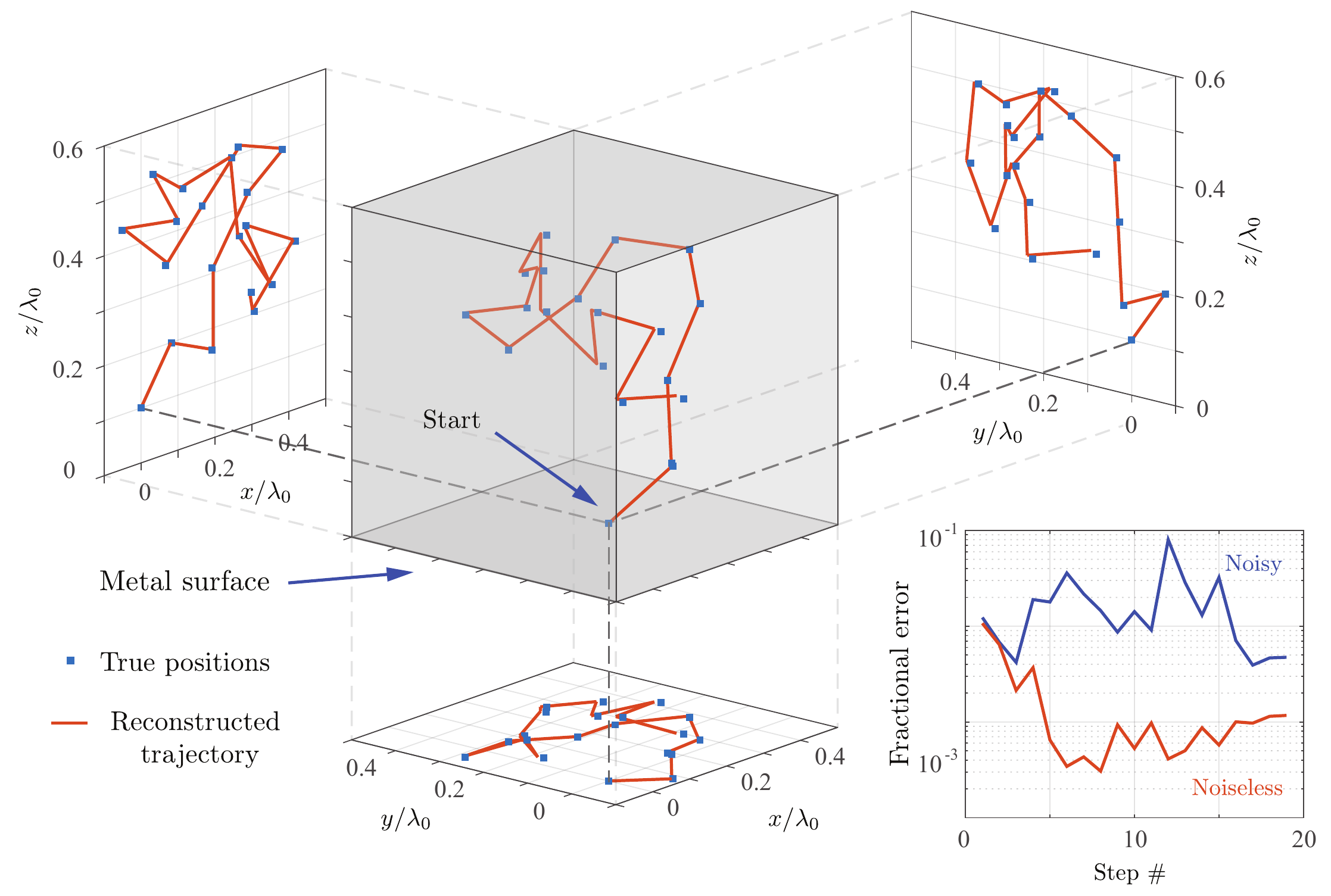}
	\caption{Reconstruction of a 3D random walk (orange line) with projections onto each coordinate plane. Blue squares show the nominal positions of the analyte particle at each frame. The initial start point of the reconstructed trajectory was set to coincide with the true position. Inset shows the fractional error ($|\delta\hat{\mathbf{r}} - \delta{\mathbf{r}} | / |\delta{\mathbf{r}} |$) in the estimated step  for each step in the trajectory for the shot noise limited (blue) and noise free (orange) case.}
	\label{fig:trajectory_reconstruction}
\end{figure*}

\subsection{Sign Ambiguity of Transverse Step}
From \eqref{eq:phase_shift} it can be seen that $\Psi(\phi;-\delta\boldsymbol{\rho},-\delta z)=-\Psi(\phi;\delta\boldsymbol{\rho},\delta z)$ meaning that, ignoring the effects of the decay, the negative of the correct step  also satisfies \eqref{eq:cost_func}. In reality, the decay of the field in the $x$ and $z$ directions ensures $-\delta\mathbf{r}$ is not a solution, however in cases where the decay is weak (e.g. small $\delta z$) the exact backward step corresponds to either a very low local minimum in the minimisation landscape, or, if noise is strong enough to obscure any decay, a global minimum. Since the decay in the $z$ direction is over a much shorter length scale than in the $x$ direction ($\kappa_w' \gg k_{sp}''$), and the phase shift depends only weakly on $\delta z$, in practice, this incorrect sign error is only seen in the estimate of the transverse step. A simple strategy to protect against possible sign ambiguities is to first run the minimisation algorithm to solve \eqref{eq:cost_func} as usual and then run a second minimisation using the retrieved step as the initial start point albeit with the transverse step reversed.  If the second minimisation converges to a lower minimum, the new step estimate is taken, otherwise the initial estimate is retained. Use of this strategy means any sign error can only remain if the  flipped transverse step truly corresponds to a lower minimum than the true step as a consequence of noise. In this case, incorrect step estimates can be resolved by applying the consistency check to both the estimate from the global minimum, and the same estimate but with the transverse step flipped, and then picking the sign of the transverse step which is most consistent with the estimates generated from overlapping sets of three frames. It should be emphasised that in the majority of cases, the decay factor ensures that there is no sign error and the true step corresponds to the global minimum.

\section{Algorithm Performance}\label{sec:results}

\begin{figure}[t]
	\centering
	\includegraphics[width=\columnwidth]{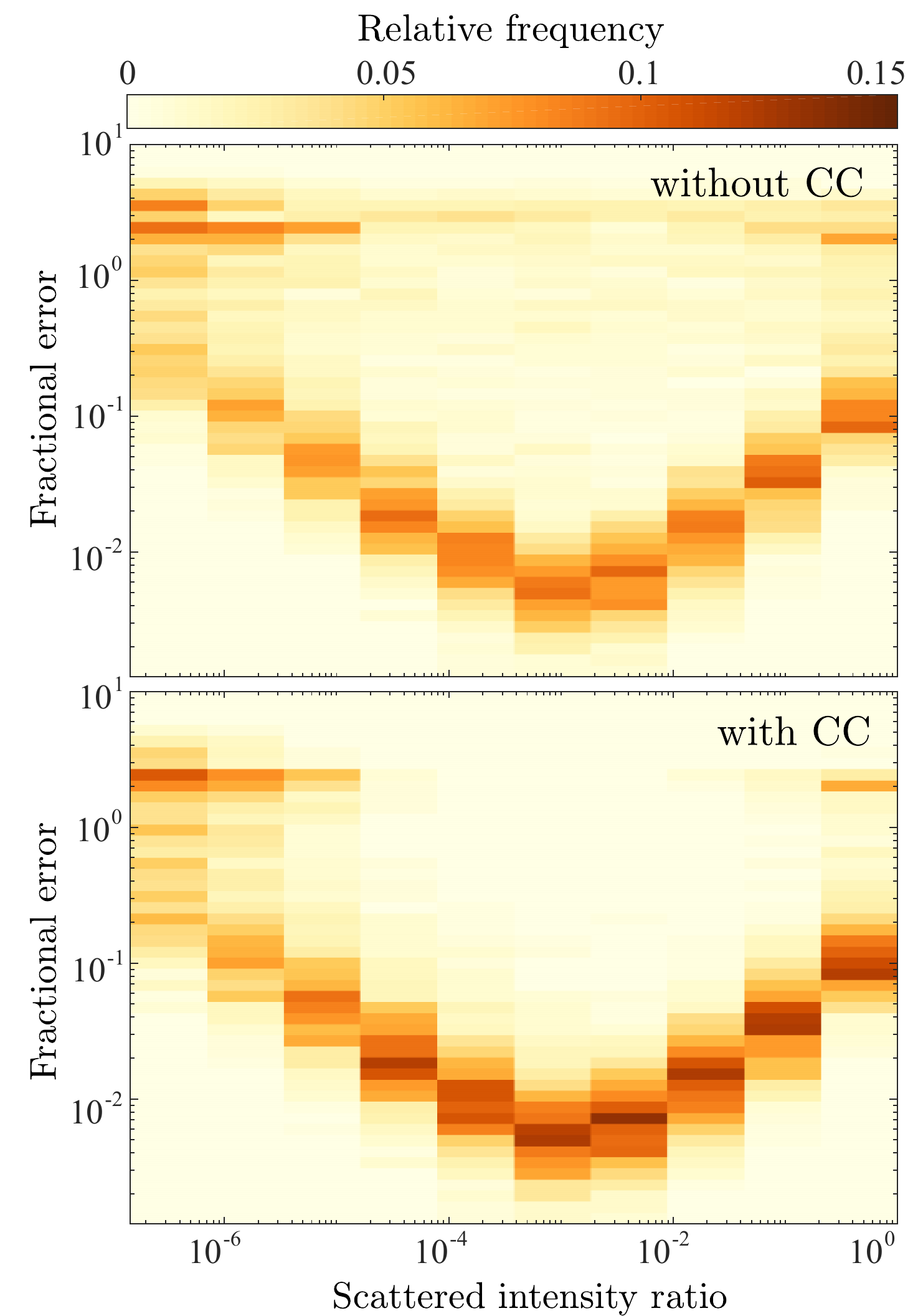}
	
	\caption{The dependence of the relative frequency of a given fractional error ($|\delta\hat{\mathbf{r}} - \delta{\mathbf{r}} | / |\delta{\mathbf{r}} |$) on the ratio of total particle scattered intensity to total background intensity, $ \sum_k\langle\widetilde{I}_s(\phi_k) \rangle/ \sum_k\langle\widetilde{I}_b(\phi_k) \rangle$, found  without (top) and with (bottom) the consistency check. The step size was fixed at $|\delta\mathbf{r}|=0.15\lambda_0$. The background was fixed to be $10^{10}$ photons around the entire ring, which consisted of 100 azimuthal pixels. Distributions shown were calculated using  1050 different realisations of background speckle and particle trajectories for each value of the particle scattered power. All intensity distribution were corrupted by a shot noise. The consistency check was applied a maximum of five times, after which the lowest minimum was taken.}
	\label{fig:error_vs_photon_number}
\end{figure}

The proposed method was tested and verified using simulated data. In particular, the value of the background field at each pixel was constructed by first generating a complex zero mean Gaussian random vector of length $N$, which was then convolved with a Gaussian smoothing function to yield a ring of speckle with finite speckle size with a desired average angular width. The chosen speckle size was found to have no significant effect on the algorithm, but was chosen to be $0.05$ radians, corresponding to observed leakage ring speckle patterns. The analyte particle was modelled as a dipole scatterer with dipole moment aligned with the incident SPP field. The dipole approximation is only valid for small particles, for which the integrand in \eqref{eq:born} does not vary significantly over the particle volume \cite{Foreman2017,Chaumet1998EvanescentApproximation}. Note, however, that the derivation of the tracking algorithm does not require this approximation to be valid.  To calculate the far field scattered dipole field  the standard semi-analytic thin film Green's tensor \cite{Novotny2012PrinciplesNano-optics} was used. The particle was assumed to undergo a 3D random walk in water (refractive index $n_w=1.33$ \cite{water_dielectric}), starting $60$~nm above a 40~nm thick gold  film (refractive index $n_m=0.28+2.93i$ \cite{dielectric_gold}) on top of a glass substrate (refractive index $n_g=1.5$). \comment{Whilst the simulated trajectories were generated assuming a random walk, the algorithm requires no such restriction and can track more complicated particle motion. Indeed, in general, the analyte particle will not undergo a free random walk due to optical forces arising from the SPP field \cite{Min2013}, convection and thermophoresis caused by heating effects \cite{GarcesChavez06}, in addition to any interaction or surface potential that the analyte particle may experience from bound receptors or surface charges \cite{Wong97}. Generating the trajectories in our simulations randomly however helps ensure an unbiased sampling of propagation directions so that algorithm performance can be more uniformly tested.} A free space wavelength of $\lambda_0=600$~nm was taken  and the random walk was generated with a fixed step size of $0.15\lambda_0$. With these physical parameters, the short range SPP wavenumber and decay constant are $k_{sp}=(1.50 + 0.04i)k_0$ and $\kappa_w=(0.68 + 0.08i)k_0$ respectively. The minimisation was performed using a multi-start conjugate gradient method, with 100 random starting points and the search space was bound to step sizes below $\lambda_0/2$.  Minimisation was performed on all sets of consecutive frames $i-1$, $i$, $i+1$, with the result of the minimisation for each set of three frames providing estimates $\delta\mathbf{\hat{r}}_{ij}$ of $\delta\mathbf{r}_{ij}$ for $j=i-1$ and $j=i+1$. A consistency check between step estimates from overlapping sets of frames (i.e. $\delta\mathbf{\hat{r}}_{i,i+1}$ obtained from frames $i-1$, $i$ and $i+1$ compared to $\delta\mathbf{\hat{r}}_{i,i+1}$ obtained from frames $i$, $i+1$ and $i+2$) was performed. If the steps were found to be inconsistent (the threshold for inconsistency was step estimates differing by more than 10\%), the minimisation procedure was repeated on the two relevant inconsistent sets of three frames, for 200 new random starting points, with new step estimates replacing the previous estimate if a lower minimum was found. The consistency check was repeated with the new estimates, and global minimisation repeated if the consistency check failed, adding 100  random starting points after each failure, up to a maximum of five times before the algorithm was terminated. For noiseless simulations, the algorithm was able to accurately reconstruct the particle trajectory, for sub-wavelength step sizes up to $\lambda_0/2$, with fractional errors below 1\%, corresponding to sub nanometre precision (see inset of Fig. \ref{fig:trajectory_reconstruction}). In addition to a randomly generated background speckle, the method was also tested on simulated speckle patterns generated by coupled dipole simulations \cite{Novotny2012PrinciplesNano-optics}, resulting in similar levels of accuracy.

It is also important to consider the performance of the tracking algorithm when applied to noisy data, since any real measurement cannot fully eliminate noise. To study the robustness of the tracking algorithm to noise, further simulations were performed assuming shot noise limited intensity measurements. Accordingly, the intensity measured in each azimuthal pixel was derived from a Poisson distributed random variable with mean and variance corresponding to the number of photons $\widetilde{N}_k = \widetilde{I}_i(\phi_k)A \Delta \tau/( \hbar\omega)$ incident on that pixel (of area $A$) during a single frame (integration time of $\Delta\tau$). Fig.~\ref{fig:error_vs_photon_number} shows the relative frequency of a given fractional error in the estimated step, defined as $\abs{\delta\hat{\mathbf{r}}_{ij}-\delta\mathbf{r}_{ij}}/\abs{\delta\mathbf{r}_{ij}}$ as a function of the ratio of the average total intensity scattered by the particle and the average total background intensity (i.e. $ \sum_{k=1}^N \langle\widetilde{I}_s(\phi_k)\rangle  / \sum_{k=1}^N \langle\widetilde{I}_b(\phi_k) \rangle$ where averages here are taken over an ensemble of 1050 different noise realisations, speckle realisations and particle trajectories). For the simulations the total number of photons in the background speckle integrated around the ring ($\widetilde{N}_b = \sum_{k=1}^N\langle\widetilde{I}_b(\phi_k)\rangle A\Delta \tau/( \hbar\omega) $) was assumed to be fixed at $10^{10}$. For small particle scattered intensity, the change in intensity as the particle moves is obscured by the shot noise, increasing the error in the step estimate. On the other hand, the reduced performance as the particle scattered intensity becomes comparable to the background intensity is due to the direct scattering term becoming significant and thus the approximations used in deriving the algorithm no longer hold. A significant reduction in the average error is seen when the consistency check is utilised. When the check is not used the algorithm can occasionally converge to a local minima corresponding to an incorrect step. An example reconstructed trajectory for shot noise limited measurements (at an intensity ratio $\sum_{k=1}^N \langle\widetilde{I}_s(\phi_k)\rangle  / \sum_{k=1}^N \langle\widetilde{I}_b(\phi_k) \rangle=10^{-3}$) is also shown in Fig.~\ref{fig:trajectory_reconstruction}. Whilst reconstruction errors can be seen, there is nevertheless generally good agreement with the nominal trajectory with errors $\lesssim 1$~nm.
\newcommand{\STAB}[1]{\begin{tabular}{@{}c@{}} #1\end{tabular}}
\begin{table*}[t!] 
	\caption{Summary of noise metrics for dark field and interference methods}
	\label{tab:snr}
	\centering
	\begin{tabular}{ |c||c|c|c|c|}
		
		\hline
		& \makecell*{Metric}&\makecell*{Exact expression}&\makecell*{Shot noise limit\\${N}_{T}\gg{N}_{d}$} &\makecell*{Dark noise limit\\${N}_{T}\ll{N}_{d}$}\\
		\hline\hline
		\multirow{2}{*}{\STAB{\rotatebox[origin=c]{90}{Dark field}}}
		&   \makecell*{SBR}  
		& \makecell*{$\frac{{N}_s}{{N}_{d}}$}  &\makecell*{$\frac{{N}_s}{{N}_{d}}\gg1$}
		&\makecell*{$\frac{{N}_s}{{N}_{d}}\ll1$}\\
		\cline{2-5}
		& \makecell*{$\textrm{SNR}$ }   &\makecell*{$\frac{{N}_s}{\sqrt{{N}_s+{N}_{d}}}$}
		& \makecell*{$\sqrt{{N}_s}$}
		& \makecell*{$\frac{{N}_s}{\sqrt{N_{d}}}$}\\
		\hline
		\multirow{2}{*}{\STAB{\rotatebox[origin=c]{90}{ Interference}}} 
		& \makecell*{SBR}
		& \makecell*{$\frac{\sqrt{{N}_s{N}_b}\abs{\cos\Phi}}{{N}_T+{N}_{d}}$} 
		&\makecell*{$\sqrt{\frac{{N}_s}{{N}_b}}\abs{\cos\Phi}\ll1$}
		&\makecell*{$\frac{\sqrt{{N}_s{N}_b}}{{N}_{d}}\abs{\cos\Phi}\ll1$}\\
		\cline{2-5}
		&  \makecell*{SNR}
		&\makecell*{$\frac{\sqrt{{N}_s{N}_b}\abs{\cos\Phi}}{\sqrt{{N}_T+{N}_{d}}}$} 
		& \makecell*{$\sqrt{{N}_s}\abs{\cos\Phi}$} 
		& \makecell*{$\frac{\sqrt{{N}_s{N}_b}}{\sqrt{{N}_{d}}}\abs{\cos\Phi}$} \\
		
		\hline
		
	\end{tabular}
\end{table*}
\section{Discussion}\label{sec:discussion}
To help illustrate the advantages of our proposed interferometric method, we now consider the signal to background ratio (SBR) and signal to noise ratio (SNR) for an interferometric measurement as compared to those found for a dark-field technique \cite{Weigel14,Chazot2020} in which there is no background. Importantly, for any given pixel  the  number of incident photons measured per integration time $N_T$ (note that we henceforth drop the tilde notation and $k$ subscript for clarity) can arise from multiple sources, such as the background, scattered light or noise. Changes in the number of scattered photons, $N_s$, however, derive from the presence (and movement) of the analyte particle and thus constitute an information carrying signal. Accordingly, in the definition of SBR and SNR,
\begin{align}
    \textrm{SBR}&=\frac{\langle S\rangle}{\langle N_0\rangle}\\
    \textrm{SNR}&=\frac{\langle S\rangle}{\sigma},
\end{align}
the `signal' can be expressed (for small $N_s$) as $S=\abs{\partial N_T/\partial N_s} N_s$, whereas $\sigma$ is the standard deviation of $N_T$, $N_0$ is the background photon count in the absence of the analyte particle and angled brackets denote averaging over noise realisations.  For a fair comparison between the inteferometric and dark field techniques, we assume that $N_s$ is the same for both methods. In the dark field case, the total number of signal photons incident on a single pixel per integration time can hence be entirely attributed to scattering from the analyte particle, so that ${N}_T={N}_s$ and the signal is simply $S_{\textrm{dark}}=N_s$ as would be expected.  On the other hand, for an interferometric method the contribution from the reference background field must also be considered.  The total number of incident photons is therefore ${N}_T={N}_s+{N}_b+2({N}_s{N}_b)^{1/2}\cos\Phi$, where $N_b$ is the number of background photons incident on the pixel per integration time and $\Phi$ is the phase difference between the background  field and the field scattered from the analyte particle. We again assume that ${N}_b\gg{N}_s$, whereby the inteferometric signal is $S_{\textrm{int}}=(N_sN_b)^{1/2}\abs{\cos\Phi}$. 

 Given the faint signals involved, two possible sources of noise that can affect both techniques are that of shot noise (arising from quantisation of light) and dark noise (arising from thermal excitation of electrons in the detector). As such the recorded number of photons for any given measurement can be modelled using a Poisson random variable, with mean  given by the number of photons incident on a given pixel per integration time, ${N}_T$, plus a signal independent mean dark count ${N}_d$. In the dark field case, the background is entirely due to dark noise, $N_0=N_d$, while the interferometric case has a background $N_0=N_d+N_b$. Using the fact that the variance of a Poisson distribution is equal to its mean, the  standard deviation of the noise follows as  $\sigma=({N}_{T}+{N}_{d})^{1/2}$. We can hence identify two limiting cases, namely the shot noise and dark noise limited regimes, for which ${N}_{T}\gg{N}_{d}$ and ${N}_{T}\ll{N}_{d}$ respectively.  Table~\ref{tab:snr} summarises the form of each metric for both dark field and interferometric measurements in each of the different noise regimes. In the case where the methods are shot noise limited, the SNR of both methods are of comparable size, with the interferometric method having a significantly reduced SBR.
The benefits of an interferometric method are seen when  ${N}_s\ll{N}_{d}\ll{N}_b$, in which the interference method will be in the shot noise limit, whilst the dark field method will be dark noise limited. This case is particularly relevant for tracking of  single biological particles, since their small scattering cross-section implies $N_s$ is small, whilst the background strength can to some extent be increased by appropriate experimental design. Regardless of whether an interference based method is shot noise or dark noise limited, we note that it achieves improved SNR over dark noise limited dark field measurement when ${N}_{d}>\sqrt{{N}_s}$, again at the cost of a worse SBR. The ability to achieve shot noise limited measurements even for small $N_s$ below the dark noise limit is a significant advantage within an interferometric measurement scheme. Analogous conclusions have been reached in regards to iSCAT \cite{Taylor2019InterferometricScattering}. 

\comment{The proposed setup here is similar to that of iSCAT using surface plasmon illumination, referred to as interferometric plasmonic microscopy (iPM) \cite{Yang2018InterferometricExosomes}, which itself is similar to conventional wide field iSCAT in reflection, but with a Kretschmann type glass-gold-sample thin film structure instead of a glass cover slip. Like these two iSCAT setups, the proposed method has a shared path between the reference and scattered field, meaning no requirement for a reference arm and eliminating the need for interferometric stability. There exist several alternative iSCAT setups, with either wide field or point scanning illumination, and imaging in either reflection or transmission \cite{Taylor2019InterferometricScattering, Lindfors04,Arbouet04,Ignatovich06,Jacobsen06}. A major difference between iSCAT and our proposed method is that the former (including iPM) requires direct imaging of the surface, and subsequent image processing for particle localisation via fitting of a point spread function, whilst our method uses Fourier plane imaging, using the entire Fourier plane leakage ring intensity profile to extract the particle shift.} There are a number of merits to our proposed setup compared to conventional iSCAT methods. Firstly,  light (both background and scattered from the analyte) is confined to the leakage radiation cone, resulting in larger intensities as compared to more diffuse scattering, therefore increasing the SNR. \comment{Directional scattering from the analyte particles near dielectric interfaces has been used in iSCAT to improve the SBR, with a partial reflector filtering out some of the reference reflected light while allowing the highly directional light scattered by analyte particles to pass \cite{Liebel2017UltrasensitiveProteins,Cole2017Label-FreeMicroscopy}, but these do not increase the SNR.} Scattered intensities (and thus the SNR) are also increased by virtue of the strong confined field of the illuminating SPP. Furthermore, since scattering into the ring is mediated by SPPs, both the background and particle scattered field are predominantly p-polarised. This (approximate) co-polarisation helps to ensure that interference contrast is maximised. Additionally, the use of a random speckle background as opposed to a well defined reference field means, rather than acting as a noise source as in iSCAT, scattering impurities play a central role in our technique since they act as the source of the background. \comment{Notably, background removal is an essential part of any iSCAT measurement \cite{Geme18}, which is typically achieved by analysing frame to frame differential  or ratiometric images. Our proposed algorithm however automatically accounts for the background reference field including all random static scatterers. Any additional sources of scattering, such as complex dielectric environments or refractive index variations arising from inhomomgeneous temperature variations, will simply contribute to the background speckle. The proposed method is hence robust against such impurities, provided they do not vary rapidly in time.} As well as eliminating the detrimental effects of speckle, our proposed technique also has the added benefit of not requiring high quality smooth metal surfaces, since roughness is the source of the background speckle. 

Finally, the validity of the assumptions, in particular the single scattering approximation, should be considered. For the physical setup considered, it is useful to consider two types of multiple scattering paths, ones in which light is multiply scattered by the analyte particle (in addition to an arbitrary number of scattering events from the surface scatterers), and those paths which involve multiple scattering events by the surface but only a single scattering event from the analyte particle. Since we are interested in weakly scattering particles, the first class of multiple scattering paths will contribute negligibly, since each additional scattering event from the particle introduces another factor of the (small) particle scattering strength. Since the particle can be treated as only singly scattering, the total field can hence be written as $\widetilde{\mathbf{E}}(\phi)=\widetilde{\mathbf{E}}_b(\phi)+\widetilde{\mathbf{E}}_s(\phi)$, where the background field is independent of the analyte particle. The second class of multiple scattering paths in general may be significant, since the metal surface features may scatter strongly and, depending on the surface, may be sufficiently close together for multiple scattering to be non-negligible. Since the algorithm does not make any assumptions on the nature of the background speckle, the effect of this multiple scattering on $\mathbf{E}_b$ is irrelevant. Additionally, the frequency shift and broadening of the SPP resonance associated with multiple scattering is not important, since, as mentioned earlier, $k_{sp}$ can be experimentally determined. Multiple scattering from the metal surface will, however, have an effect on $\mathbf{E}_s$, since the local field the particle experiences will include a surface scattered contribution, in addition to the incident SPP. This means that the incident field, $\mathbf{E}_0(\mathbf{r})$ in the Born approximation (\eqref{eq:born}) should be replaced by $\mathbf{E}(\mathbf{r'})=\mathbf{E}_0(\mathbf{r'})+\mathbf{E}_b(\mathbf{r'})$, where $\mathbf{E}_b(\mathbf{r'})$ is the field scattered from the surface to the position of the particle (i.e. the background speckle field, but evaluated at the position of the analyte particle instead of in the far field). Propagating this additional term through the derivation, one finds the same phase/decay factors from shifting the Green's tensor, but there are also unknown phase/amplitude changes in $\mathbf{E}_b(\mathbf{r})$, corresponding to a near field speckle distribution. \comment{The presence of SPP hotspots, regions in which $\mathbf{E}_b(\mathbf{r})$ is large, can lead to enhanced scattering from the analyte particle owing to the larger induced dipole. This enhancement in $\mathbf{E}_s$ is a multiple scattering effect, since it involves the scattering from the analyte particle of light already scattered by the surface.} Therefore, in order for the method to work, the surface scattered field on the particle has to be weak compared to the incident field. \comment{This can be achieved when the root mean square height deviation and transverse correlation length are sufficiently small relative to the wavelength \cite{born_approx_validity}.}  

\section{Conclusion}\label{sec:conclusion}
In this work, we have proposed a novel interferometric single particle tracking technique. The reference field used originates from random scattering of surface plasmons propagating along the surface of a rough metal film, which leak into the far field to form a speckle ring. Using a Green's tensor analysis we have derived an analytic expression describing how the interference between this leakage speckle and the field scattered from an analyte particle changes upon translation of the particle, which in turn forms the basis of our tracking algorithm. It is important to note that in our analysis no assumptions were made about the form of the background field or indeed the scattered field, save the validity of the Born approximation. As a result, the proposed method is applicable to a wider variety of situations than that considered here, since all that is required is knowledge of the appropriate phase and decay functions ($\Psi$ and $\Lambda$) for the given illumination/detection geometry. 

To verify our tracking algorithm we performed a series of Monte Carlo simulations using noisy simulated data, which  demonstrated a sub-nanometre reconstruction  accuracy in optimal conditions. 
Consideration of the number of unknowns for any given series of measurement frames highlighted that reconstruction of a particle trajectory requires simultaneous analysis of at least three different speckle images at a time. This however allows consistency checks to be applied to a reconstructed trajectory hence improving overall reconstruction quality. 

Finally, we have discussed the advantages of our surface plasmon interferometric technique in comparison to competing techniques. In particular, we have shown that, for weakly scattering analyte particles, our  technique benefits from improved SNR compared to a dark field technique at the cost of a reduced SBR. Moreover, given the random nature of the reference field used, it is also more robust to static scattering impurities which can afflict other interferometric techniques, such as iSCAT.  Given the confined nature of surface plasmons, if the analyte particle moves too far away from the metal surface, reconstruction will naturally no longer be possible. As such our proposed technique is more suitable for analysis of surface based phenomena and processes, for example study of screening/trapping potentials, transport through membrane layers, motility of bound molecular machines or binding kinetics.
 \comment{The main challenge in experimental realisation is ensuring the background scattered light is sufficiently strong to allow shot noise limited measurements, whilst still being able to detect the small intensity shifts against this background. Recently published work, in which scattering of surface plasmons has been used to image proteins and monitor their binding kinetics \cite{Zhang20}, demonstrate that such measurements are achievable.} The wide use of surface plasmon based technology, such as SPR sensors, and the relaxed fabrication tolerances, furthermore implies that proposed method could be implemented simply and cheaply and thus become a powerful tool for study of biological processes at the single particle level.

\end{document}